\documentclass[aps,prb,showpacs,twocolumn,superscriptaddress]{revtex4-1}
\usepackage{epsfig,amsmath,amssymb,color}
\newcommand{\be}{\begin{equation}}
\newcommand{\ee}{\end{equation}}

\newcommand{\ra}{\rangle}
\newcommand{\la}{\langle}
\newcommand{\bit}{\begin{itemize}}
\newcommand{\eit}{\end{itemize}}
\newcommand{\bea}{\begin{eqnarray}}
\newcommand{\eea}{\end{eqnarray}}

\usepackage{epsfig}

\begin{document}
\title
{The ground-state phase diagram of the XXZ spin-$s$ kagome antiferromagnet: A coupled-cluster study}

\author
{ O. G\"otze and J. Richter\\
\small{Institut f\"ur Theoretische Physik, Universit\"at Magdeburg,
39016 Magdeburg, Germany}\\
}

\begin{abstract}
We use the coupled cluster method to high orders of approximation
in order to calculate the 
ground-state phase diagram of the XXZ spin-$s$ kagome antiferromagnet with
easy-plane anisotropy, i.e. the anisotropy parameter $\Delta$ varies between
$\Delta=1$ (isotropic Heisenberg model) and $\Delta=0$ ($XY$ model).
We find that for the extreme quantum case $s=1/2$ the ground state is
magnetically disordered  in the entire region  $0 \le \Delta \le 1$.
For $s=1$ the ground state is disordered for $0.818 < \Delta \le 1$, it
exhibits 
$\sqrt{3}\times\sqrt{3}$ magnetic long-range order for  
$0.281 < \Delta <0.818$, and $q=0$ magnetic
long-range order for $0 \le \Delta < 0.281$.
We confirm the recent result of Chernyshev and
Zhitomirsky ({\it  Phys. Rev. Lett. {\bf 113}, 237202 (2014)})
that the selection of the ground state by quantum fluctuations is different
for small $\Delta$ ($XY$ limit) and for $\Delta$ close to one (Heisenberg limit),
i.e.,
$q=0$ magnetic order is favored over $\sqrt{3}\times\sqrt{3}$ for
$0\le \Delta <\Delta_c$ and vice versa for $\Delta_c < \Delta \le 1$. We calculate
$\Delta_c$ as a function of the spin quantum number $s$. 
\end{abstract}
\maketitle

{\it Introduction.}
The investigation of the ground state (GS) of the quantum antiferromagnet
 on the kagome lattice is one of the most
challenging problems in the field of frustrated quantum magnetism.
Over many years numerous theoretical methods 
has been applied to understand the GS properties of the kagome
antiferromagnet (KAFM),
see,
e.g.,~Refs.~\onlinecite{Elser1990,Harris1992,chub92,sachdev1992,Reimers1993,
henley1995,Waldtmann1998,Capponi2004,Singh2007,cepas2008,Sindzingre2009,Bishop2010,Evenbly2010,Yan2011,lauchli2011,
lee2011,nakano2011,goetze2011,schollwoeck2012,
Li2012,becca2013,bruce2014,Ioannis2013,zhito_XXZ_2014,XXZ_s12_2014,XXZ_s12_2014_a,cepas2011,Lauchli_s1_2014,Weichselbaum_s1_2014,satoshi_s1_2014,Weichselbaum_s1_2014a} 
and references
therein.

While it became clear very early that GS magnetic long-range order (LRO) is
absent for the  $s=1/2$ Heisenberg KAFM, there was a
longstanding debate on the nature of the non-magnetic quantum GS.  
Recent large-scale numerics\cite{Yan2011,lauchli2011,schollwoeck2012}
provide strong arguments for a gapped Z2 topological spin-liquid GS for spin
quantum number $s=1/2$.

Other recent investigations have been focused on higher spin
$s>1/2$\cite{cepas2011,goetze2011,zhito_XXZ_2014,Lauchli_s1_2014,Weichselbaum_s1_2014,satoshi_s1_2014,Weichselbaum_s1_2014a}
and also on the anisotropic
XXZ-model.\cite{zhito_XXZ_2014,cepas2008,XXZ_s12_2014}
Both modifications have relevance for the experimental research, see, e.g.,
Refs.~\onlinecite{jaros1,jaros2,s32_a,s32_b,s1,DM1,DM2,DM3,Tanaka2014}. 
Moreover, anisotropic spin models are of great interest with respect to 
engineering models of quantum magnetism on optical lattices, see, e.g.,
Refs.~\onlinecite{isakov2006,bloch2008,struck2013}.
Since higher spin quantum numbers as well as spin anisotropy, in general,  lead to
a reduction of quantum fluctuations, see, e.g.,
Refs.~\onlinecite{trian_s1,j1j2_s1_ccm,trian_s1_ccm,trian_s1_ED,trian_XY,j1j2_XXZ_ccm,j1j2_honey_XY},
GS magnetic LRO
for the KAFM might be facilitated.
However, for the isotropic $s=1$ Heisenberg KAFM there is clear evidence
that there is no magnetic
LRO.\cite{goetze2011,Lauchli_s1_2014,Weichselbaum_s1_2014,satoshi_s1_2014,Weichselbaum_s1_2014a}
(Note, however, that in Ref.~\onlinecite{cepas2011} $\sqrt{3}\times\sqrt{3}$ GS LRO for integer spin quantum numbers including
$s=1$ was reported.) 
For $s>1$ several approaches lead to indications for $\sqrt{3}\times\sqrt{3}$ GS
LRO.\cite{chub92,sachdev1992,henley1995,goetze2011,cepas2011,zhito_XXZ_2014}  
On the other hand, recent density matrix group (DMRG)
calculations\cite{XXZ_s12_2014} have demonstrated that  for $s=1/2$
 the $XXZ$ KAFM
remains in
a magnetically disordered GS for the entire range of the anisotropy
parameter $\Delta$ between the $XY$ point  ($\Delta=0$) and the isotropic
Heisenberg point  ($\Delta=1$).

Motivated by the recent Letter of Chernyshev and
Zhitomirsky\cite{zhito_XXZ_2014} we use the coupled cluster method (CCM) in
high orders of approximation
to calculate the $s-\Delta$ GS  phase diagram of the  spin-$s$ XXZ
KAFM with easy-plane anisotropy.
The corresponding Hamiltonian is 
\be
H = \sum_{[i,j]} \left( s^x_is^x_j + s^y_is^y_j + \Delta s^z_is^z_j \right )
\; , \; 0\le \Delta \le 1 ,
\label{H}
\ee
where the sum runs over all nearest-neighbor  pairs.
The CCM is a very
general {\it ab initio}  many-body technique
that has been successfully applied to strongly frustrated quantum
magnets, see, e.g.,
Refs.~\onlinecite{Bishop2010,goetze2011,Li2012,j1j2_XXZ_ccm,j1j2_s1_ccm,trian_s1_ccm,j1j2_honey_XY,ferri2001_ccm,Schm:2006,
darradi08,bishop08,farnell09,richter2010,farnell11,reuther2011,bishop2013a,jiang2014}.
In particular, in Ref.~\onlinecite{goetze2011} it has been demonstrated
that the
CCM GS energy for the $s=1/2$ isotropic Heisenberg 
KAFM
is close to best available DMRG
results.\cite{Yan2011,schollwoeck2012}

{\it The coupled cluster method (CCM).}
For the sake of brevity we illustrate here only some relevant features of
the CCM. At that we follow strictly the lines given in
Ref.~\onlinecite{goetze2011}, where the CCM was applied to the isotropic
spin-$s$ Heisenberg KAFM. 
For more general information on the methodology of the CCM, see, e.g.,
Refs.~\onlinecite{zeng98,bishop98a,bishop99,bishop00,farnell02,bishop04,stiffness_ccm}.
We first mention that the CCM approach yields results directly in 
the thermodynamic limit $N\to\infty$, 
where $N$ is the number of lattice sites.

First we choose 
 a normalized reference  state
$|\Phi\rangle$ that is typically a classical GS of the model.  
From a quasi-classical point of view  
the  coplanar $\sqrt{3}\times\sqrt{3}$ and
$q=0$ states are favored candidates among the massively degenerate manifold
of classical ground states (see, e.g.,
Refs.~\onlinecite{Harris1992,Reimers1993,chub92,henley1995}).
Consequently, we use both states as reference  states.
Then we perform a rotation of the local axes of each of 
the spins such that all spins in the reference state align along the
negative $z$ axis.\cite{rotate}  
In this new set of local spin coordinates 
we define a complete set of 
mutually commuting multispin
creation operators $C_I^+ \equiv (C^{-}_{I})^{\dagger}$ related to this reference
state:
\begin{equation}
\label{set1} |{\Phi}\ra = |\downarrow\downarrow\downarrow\cdots\rangle ; \mbox{ }
C_I^+ 
= { s}_{n}^+ \, , \, { s}_{n}^+{ s}_{m}^+ \, , \, { s}_{n}^+{ s}_{m}^+{
s}_{k}^+ \, , \, \ldots \; ,
\end{equation}
where $s^{+}_{n} \equiv s^{x}_{n} + is^{y}_{n}$. In Eq.~(\ref{set1}) the
components of the spin operators are defined 
in the local rotated coordinate frames, and the indices $n,m,k,\ldots$ denote arbitrary lattice
sites, where each site index in each configuration index $I$ in
Eq.~({\ref{set1}) can be repeated up to a maximum of $2s$ times.
With the set $\{|\Phi\rangle, C_I^+\}$ thus defined, the CCM parametrizations of 
the ket 
and bra GS eigenvectors
$|\Psi\ra$ 
and $\la\tilde{\Psi}|$ 
of the spin system 
are given  by
\begin{eqnarray}
\label{eq5} 
|\Psi\ra=e^S|\Phi\ra \; , \mbox{ } S=\sum_{I\neq 0}a_IC_I^+ \; \\
\label{eq5b}
\la \tilde{ \Psi}|=\la \Phi |\tilde{S}e^{-S} \; , \mbox{ } \tilde{S}=1+
\sum_{I\neq 0}\tilde{a}_IC_I^{-} \; .
\end{eqnarray}
The correlation coefficients,
$a_I$ and $\tilde{a}_I$, 
contained  in the CCM correlation operators, $S$ and $\tilde{S}$,
are determined by the CCM ket-state 
and bra-state
equations
\begin{eqnarray}
\label{eq6}
\langle\Phi|C_I^-e^{-S}He^S|\Phi\rangle = 0 \;\; ; \; \forall I\neq 0  \\ 
\label{eq6a}\langle\Phi|{\tilde S}e^{-S}[H, C_I^+]e^S|\Phi\rangle = 0 \; \; ; \; \forall
I\neq 0.
\end{eqnarray}
Equations (\ref{eq6}) and (\ref{eq6a}) are fully equivalent to the GS Schr\"{o}dinger equations for the ket and bra states.  
Each ket-state 
or bra-state 
equation belongs to a certain configuration index $I$,
i.e., it corresponds to a certain set (configuration) of lattice sites
$n,m,k,\dots\;$, as in Eq.~(\ref{set1}).
Using the Schr\"odinger equation, $H|\Psi\ra=E_0|\Psi\ra$, we can now write
the GS energy as $E_0=\la\Phi|e^{-S}He^S|\Phi\ra$.
The magnetic order parameter (sublattice magnetization) is given
by $ M = -\frac{1}{N} \sum_{i=1}^N \la\tilde\Psi|{ s}_i^z|\Psi\ra$, where
${s}_i^z$
is expressed in the transformed coordinate system, and $N(\rightarrow \infty)$ is the number of lattice sites. 

For the quantum many-body system under consideration
we have to use an appropriate approximation 
scheme in order to truncate the expansions of $S$ 
and $\tilde S$ 
in  Eqs.~(\ref{eq5}) and (\ref{eq5b}).
For that we use the well established SUB$n$-$n$ approximation scheme, cf., e.g.,
Refs.~\onlinecite{Bishop2010,goetze2011,Li2012,j1j2_XXZ_ccm,j1j2_s1_ccm,trian_s1_ccm,j1j2_honey_XY,ferri2001_ccm,Schm:2006,darradi08,bishop08,farnell09,
richter2010,farnell11,reuther2011,bishop2013a,jiang2014}.
 In the SUB$n$-$n$ scheme we include
no more than $n$ spin flips spanning a range of no more than
$n$ contiguous lattice sites.\cite{SUBn-n}
Using an efficient 
parallelized CCM code \cite{cccm} we are able to solve the CCM equations up
to SUB10-10 for $s=1/2$ 
and up
to SUB8-8 for $s>1/2$.
We have calculated the GS energy per spin $e_0=E_0/N$ and the magnetic order parameter    
$M$ for spin  quantum numbers $s=1/2,1,\ldots,9/2,5$.
The maximum number of ket-state equations which we have to take into account
is  416193 for $s=5$.
Following Ref.~\onlinecite{goetze2011} we extrapolate the `raw' SUB$n$-$n$
data to the limit $n \to \infty$
using $n=4,5,\ldots,10$ ($n=4,5,\ldots,8$) for $s=1/2$ ($s>1/2$).
For that  
we use
the well-tested extrapolation
ans\"atze\cite{Bishop2010,goetze2011,Li2012,ferri2001_ccm,j1j2_s1_ccm,trian_s1_ccm,j1j2_honey_XY,Schm:2006,darradi08,bishop08,farnell09,
richter2010,farnell11}
$e_0(n) = a_0 + a_1(1/n)^2 + a_2(1/n)^4$ and 
$M(n)=b_0+b_1(1/n)^{1/2}+b_2(1/n)^{3/2}$.

\begin{figure}[ht]
\begin{center}
\epsfig{file=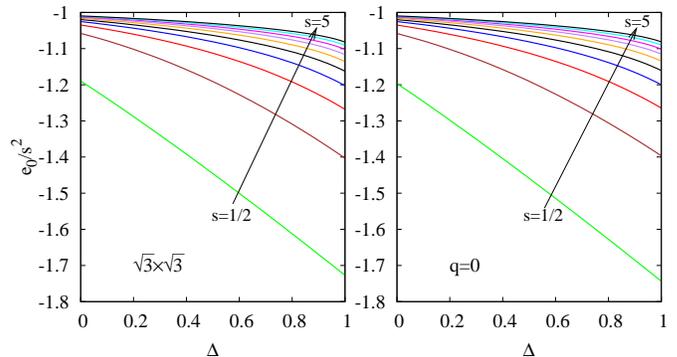,scale=0.56,angle=0.0}
\end{center}
\caption{The extrapolated CCM GS energy per spin $e_0 \vert_{n \to \infty}$ using the
$\sqrt{3}\times\sqrt{3}$ reference state (left) and the $q=0$ reference state
(right) as a function of the
anisotropy parameter $\Delta$
for spin quantum numbers $s=1/2,1,3/2,\ldots,9/2,5$. 
}
\label{fig1}
\end{figure}

\begin{figure}[ht]
\begin{center}
\epsfig{file=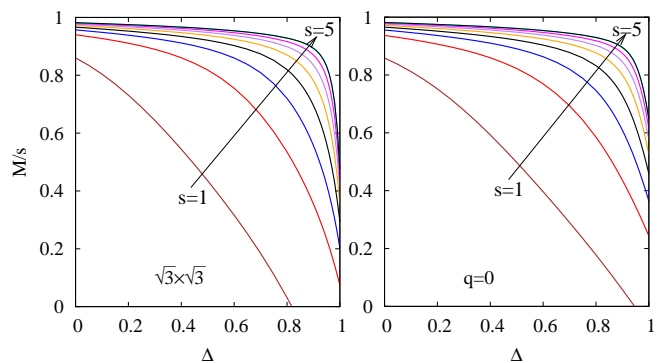,scale=0.56,angle=0.0}
\end{center}
\caption{The extrapolated CCM GS sublattice magnetization $M\vert_{n \to
\infty}$ using the
$\sqrt{3}\times\sqrt{3}$ reference state (left) and the $q=0$ reference state
(right) as a function of the
anisotropy parameter $\Delta$
for spin quantum numbers $s=1,3/2,\ldots,9/2,5$. 
Note that for $s=1/2$ no data are shown, since there is no magnetic LRO in the entire region $0\le \Delta \le
1$.
For $s=1$ there are finite regions of disorder $\Delta^* \le \Delta \le1$ for both reference
states.
}
\label{fig2}
\end{figure}

\begin{figure}[ht]
\begin{center}
\epsfig{file=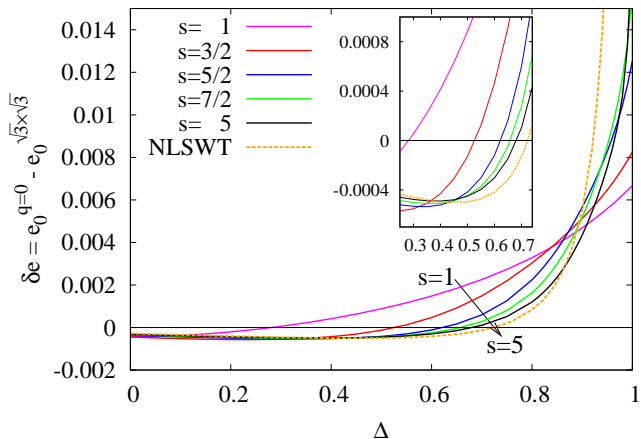,scale=0.70,angle=0.0}
\end{center}
\caption{Difference 
$\delta e = e_0^{q=0} - e_0^{\sqrt{3} \times \sqrt{3}}$
of the  extrapolated GS energies of the 
 $\sqrt{3}\times\sqrt{3}$ and the $q=0$ states 
as a function of the
anisotropy parameter $\Delta$
for spin quantum numbers $s=1,3/2,5/2,7/2$ and $5$.
For comparison we also show the corresponding large-$s$ results of
Ref.~\onlinecite{zhito_XXZ_2014} (labeled by 'NLSWT') obtained by non-linear SWT.    
The inset shows an enlarged scale of that region of $\Delta$, where $\delta e$ changes its
sign.   
}
\label{fig3}
\end{figure}

\begin{figure}[ht]
\begin{center}
\epsfig{file=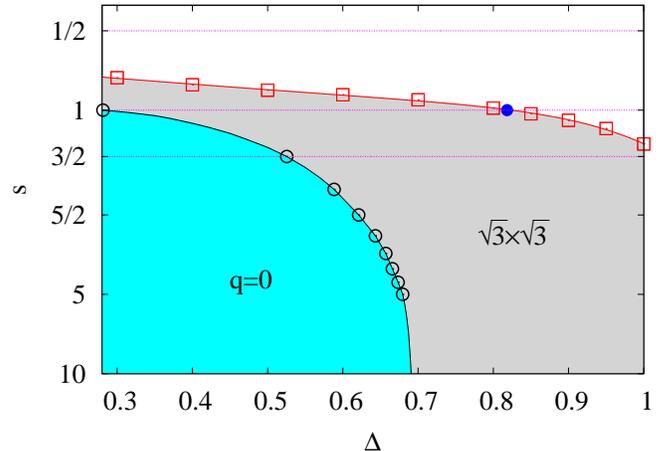,scale=0.73,angle=0.0}
\end{center}
\caption{GS $s-\Delta$ phase diagram
of the XXZ spin-$s$ kagome antiferromagnet.
The open circles connected by the black solid line show the critical anisotropy $\Delta_c(s)$, where the 
GS selection changes   
between the $\sqrt{3}\times\sqrt{3}$ state and $q=0$ state.
The upper red line connecting the open squares is the fictional transition line 
between GS disorder and GS LRO, which is obtained by considering the spin
quantum number $s$ as a continuous quantity. 
The blue point indicates the transition point $\Delta^*=0.818$ between the disordered
ground state and
the ground state with $\sqrt{3}\times\sqrt{3}$ LRO for $s=1$ obtained directly from
the $M(\Delta)$ curve shown in Fig.~\ref{fig2}. 
}
\label{fig4}
\end{figure}

{\it Results.}
As already mentioned in the introduction, we want to calculate the 
GS $s-\Delta$ phase diagram.
Such a
phase diagram has been very
recently presented  by Chernyshev and
Zhitomirsky\cite{zhito_XXZ_2014} using 
nonlinear spin-wave and real-space perturbation theories.
To have the necessary information available for the comparison  of Chernyshev's and
Zhitomirsky's results  with our CCM results
reported below,  
let us first briefly report the main findings of Ref.~\onlinecite{zhito_XXZ_2014}. 
It is well known\cite{Harris1992,chub92,henley1995,goetze2011} that in the large-$s$ limit
for $\Delta=1$ quantum
fluctuations select the
$\sqrt{3}\times\sqrt{3}$ state. In  Ref.~\onlinecite{zhito_XXZ_2014} it was
found  that this GS selection is preserved for weak
easy-plane anisotropy, i.e. for $\Delta_c < \Delta
\le 1$.
However, the main and unexpected result of Ref.~\onlinecite{zhito_XXZ_2014} is the
selection of the $q=0$ GS  for smaller values of $\Delta$ down to the $XY$
point, i.e. for $0 \le \Delta < \Delta_c$.
This finding   
is contrary to  the 
selection trend by thermal fluctuations for the classical KAFM, where for $\Delta=1$ and for
$\Delta=0$ the $\sqrt{3}\times\sqrt{3}$ state is asymptotically
selected.\cite{Harris1992,huse1992,henley2009,Reimers1993,korshunov}
Hence, for the $XY$ antiferromagnet on the kagome lattice we are faced with an example
that
quantum and thermal fluctuations may act very differently.
The term  in the nonlinear spin-wave theory (SWT) responsible for the GS selection is  of order ${\cal
O}(1)$, and, therefore, the critical anisotropy  $\Delta_c=0.722 35$ is found
to be independent of the spin quantum number $s$ within this
approximation.\cite{zhito_XXZ_2014}
The real-space perturbation theory provides insight in the mechanism of the
quantum selection of the ground state:  Some relevant seventh-order
processes  change their sign as varying $\Delta$.\cite{zhito_XXZ_2014} 
The magnetic order parameter calculated in harmonic approximation shows
a clear trend to magnetic LRO as lowering $\Delta$.
However, this trend is substantially overestimated by SWT, 
since already for
the extreme quantum case, $s=1/2$, the GS is found to be magnetically long-range ordered for $\Delta < 0.95$.
On the other
hand, for the  isotropic Heisenberg point it is well-known that the linear
SWT yields a vanishing  order parameter for any value of
$s$.\cite{Elser1990,zhito_XXZ_2014}
Both findings  for the order parameter are in contradiction to 
recent results obtained by large-scale numerical approaches.\cite{XXZ_s12_2014,goetze2011}

We now discuss our CCM results.
In Fig.~\ref{fig1} and  Fig.~\ref{fig2} we show the GS energy per spin $e_0$ and the magnetic order parameter
$M$  extrapolated to $n \to
\infty$ 
for both the $\sqrt{3}\times\sqrt{3}$  and the $q=0$
reference states 
 as a function of $\Delta$ for spin quantum numbers up to $s=5$.
In general, the $\sqrt{3}\times\sqrt{3}$  and the $q=0$ cases behave very
 similar.
The GS energy for $s=1/2$ is almost linearly growing with decreasing of
$\Delta$. For larger $s$ the $e_0(\Delta)$-curve noticeably deviates from
linearity, particularly  near $\Delta=1$.
This trend that the influence of the anisotropy $\Delta$ becomes
exceedingly large  near the isotropic Heisenberg limit at 
$\Delta=1$ is more pronounced for the order parameter, see Fig.~\ref{fig2}.
For spin quantum numbers $s \ge 3/2$ there is a drastic downturn in
the $M(\Delta)$-curve as approaching $\Delta=1$ and  there is a strong
increase in the slope $(dM/d\Delta)\vert_{\Delta=1}$ with growing $s$.
Note that a special behavior for $\Delta\to 1$ is also present  within the spin wave
approach, where the $1/s$ corrections diverge for $\Delta \to 1$.\cite{zhito_XXZ_2014}
The cases $s=1/2$ and $s=1$ are different from the cases $s>1$.  For $s=1/2$
our CCM approach leads to a disordered GS in the entire 
region
of the anisotropy
parameter $0\le \Delta \le 1$ in accordance with recent DMRG
calculations.\cite{XXZ_s12_2014}  
For $s=1$ we find a finite region of disorder, $\Delta^* \le \Delta \le 1$, for both reference
states, where $\Delta^*=0.818$ ($\Delta^*=0.945$) for the  $\sqrt{3}\times\sqrt{3}$
($q=0$) reference state. Note, however, that for anisotropies   around $\Delta^*$ the CCM GS
energy for the $\sqrt{3}\times\sqrt{3}$
reference state is lower than that for the $q=0$
reference state, see below. 
We mention also that a table of values of the  GS energy and the order parameter
for $\Delta=1$ and for spin quantum numbers up to $s=3$ can be found in
Ref.~\onlinecite{goetze2011}.  
We present corresponding values for the $XY$ limit ($\Delta=0$) in Table~\ref{table1}
of the
present paper.

Next we discuss the GS selection. For that we consider the energy difference 
$\delta e = e_0^{q=0} - e_0^{\sqrt{3} \times \sqrt{3}}$, see Fig.~\ref{fig3}.
The main common feature of the curves shown in Fig.~\ref{fig3} is the change of the
sign of $\delta e$, i.e., in accord with Ref.~\onlinecite{zhito_XXZ_2014} we find a change in the GS
selection from the $\sqrt{3}\times\sqrt{3}$ state to the $q=0$ state  at a critical value
$\Delta_c$ as varying
the anisotropy from $\Delta=1$ to $\Delta=0$.
The magnitude  of $\delta e$ is small, in particular for smaller $\Delta$.
For larger $s$ it agrees very well with
the SWT data of Ref.~\onlinecite{zhito_XXZ_2014} up to $\Delta \sim 0.9$.
The stronger deviation for $\Delta$ close to one can be attributed 
to the divergence of the $1/s$ corrections for $\Delta \to
1$.\cite{zhito_XXZ_2014}

\begin{table}[htb]
    \centering
\caption{Extrapolated CCM results for GS energy per spin, $e_0 \vert_{n \to
 \infty}$,
 and the GS sublattice magnetization, $M\vert_{n \to
\infty}$, using the
$\sqrt{3}\times\sqrt{3}$ and the $q=0$ reference
states for $\Delta=0$ ($XY$ model). Note that  
corresponding tables for  $\Delta=1$ (isotropic Heisenberg  model) can be
found in Ref.~\onlinecite{goetze2011}.
}
    \begin{tabular}{lrrrr}\hline\hline
    \parbox[0pt][1.5em][c]{0cm}{}        & \multicolumn{2}{c}{$\sqrt{3}\times\sqrt{3}$} & \multicolumn{2}{c}{$q=0$} \\\hline
      &\; $e_0/s^2$ &\; $M/s$&\; $e_0/s^2$ &\; $M/s$\\ \hline
    $s=1/2$           &\; -1.1896 &\;  $<0$ &\;  -1.1968 &\;  $<0$    \\
    $s=1$           &\;-1.0578 &\;  0.8602 &\;  -1.0583 &\;  0.8589    \\
    $s=3/2$           &\;-1.0347 &\;  0.9402 &\;  -1.0349 &\;  0.9368    \\
     $s=2$           &\;-1.0252 &\;  0.9570 &\;  -1.0253 &\;  0.9556    \\
     $s=5/2$           &\;-1.0198 &\;  0.9664 &\;  -1.0199 &\;  0.9656    \\
     $s=3$           &\;-1.0163 &\;  0.9723 &\;  -1.0164 &\;  0.9719    \\
    $s \to \infty $            &\;-1 &\;  1 &\;-1 &\; 1           \\\hline\hline
    \end{tabular}
\label{table1}
\end{table}

As already mentioned above, the large-$s$  spin-wave approach\cite{zhito_XXZ_2014} yields  a critical
value $\Delta_c$ that is independent of the
 spin quantum number $s$.
However, $\Delta_c$ certainly depends on $s$.
Chernyshev and
Zhitomirsky\cite{zhito_XXZ_2014} 
suggested that $\Delta_c$ may increase for smaller
spins from the large-$s$ value $\Delta_c=0.72235$ of
$s \to \infty$.
Our CCM approach yields directly $\Delta_c$ as a function of
$s$, see the black open circles in Fig.~\ref{fig4}.
Contrary to the conjecture of Chernyshev and
Zhitomirsky, see Fig.~4b in Ref.~\onlinecite{zhito_XXZ_2014},
we find that $\Delta_c$ increases for larger $s$.
The smallest value, $\Delta_c=0.281$, is found for
$s=1$. Applying $g(x)=a + b x+ cx^2$, $x=1/s$, to extrapolate the CCM data of the critical anisotropy to $s\to\infty$
yields  $\lim_{s \to \infty}\Delta_c=0.727$, where we have used 
$s=3,7/2,4,9/2,5$ for the extrapolation.
This CCM estimate of the large-$s$ limit of $\Delta_c$ is in excellent
agreement with  the large-$s$  spin-wave result.\cite{zhito_XXZ_2014} 

To get the full relationship to the  GS $s-\Delta$  phase diagram given in
Fig.~4 of
Ref.~\onlinecite{zhito_XXZ_2014} 
we may consider the spin quantum number
$s$ as a continuous variable. We determine that (fictional) value of $s$,
for which the GS becomes magnetically disordered. For that we fit the data for
the order
parameter $M/s$ as function of $s$ and $\Delta$ 
by the fitting function
$f(s,\Delta)=b_0(\Delta)-b_1(\Delta)s^{-1/2}-b_2(\Delta)s^{-1}-b_3(\Delta)s^{-3/2}-b_4(\Delta)s^{-2}$,
where $s=1/2$ is excluded from the fit.
From $f(s_c,\Delta)=0$
we obtain the corresponding phase boundary $s_c(\Delta)$  between magnetically
disordered and ordered GS phases. This phase boundary  is shown by the red solid line in
Fig.~\ref{fig4}.  
The  resulting value of $s_c$ for  $\Delta =1$  ($\Delta =0$) is $s_c \sim 1.34$
($s_c \sim 0.537$). 

As already mentioned above for $s=1/2$ the GS is always magnetically
disordered. For $s=1$ there are three GS phases: the disordered
state for $0.818 < \Delta \le 1$, the ordered
$\sqrt{3}\times\sqrt{3}$ state for $0.281 < \Delta < 0.818$, 
and the ordered
$q=0$ state for $0 \le \Delta < 0.281$. For spin quantum numbers $s>1$ there
are two magnetically ordered GS phases, where the phase boundary is given by
the black solid line in Fig.~\ref{fig4}.

{\it Summary.}
We summarize our findings by comparing our 
GS $s-\Delta$ phase diagram (Fig.~\ref{fig4})
with that of
Ref.~\onlinecite{zhito_XXZ_2014}.
Most importantly, we get the same trend as Ref.~\onlinecite{zhito_XXZ_2014},
namely the GS selection changes from the $\sqrt{3}\times\sqrt{3}$ state to the $q=0$
state at a critical $\Delta_c$ as varying the anisotropy parameter $\Delta$
from $\Delta=1$ to $\Delta=0$. 
The energy difference 
$\delta e$ between the $\sqrt{3}\times\sqrt{3}$ and the $q=0$
state calculated by the CCM, e.g. for $s=5$, is in excellent agreement with
the large-$s$ SWT data of Ref.~\onlinecite{zhito_XXZ_2014}.  
Moreover, we find also in accordance  with
Ref.~\onlinecite{zhito_XXZ_2014} that the region of disorder in the GS
$s-\Delta$ phase diagram grows with growing $\Delta$.

Since our CCM approach is not limited to large $s$, we can overcome some
limitations of the SWT of Ref.~\onlinecite{zhito_XXZ_2014}:
The CCM result for the critical $\Delta_c$ depends on $s$ by contrast to the  $s$-independent
SWT value. For $s=1/2$ the SWT yields GS magnetic LRO for $\Delta \lesssim
0.95$, whereas the CCM (in agreement with DMRG results\cite{XXZ_s12_2014})  yields always a disordered
GS.

Knowing these limitations of the SWT, Chernyshev and
Zhitomirsky\cite{zhito_XXZ_2014} proposed a tentative GS $s-\Delta$ phase
diagram, see Fig.~4b of Ref.~\onlinecite{zhito_XXZ_2014}.
However, our phase diagram determined from the CCM results differs
from that conjectured by Chernyshev and
Zhitomirsky.
We find that  $\Delta_c$
increases with increasing $s$, whereas Chernyshev and
Zhitomirsky speculate that $\Delta_c$ may decrease for larger $s$. Hence our
results indicate that the
region of $\sqrt{3}\times\sqrt{3}$ 
 GS LRO is much larger than that proposed in
Ref.~\onlinecite{zhito_XXZ_2014}.  
As a consequence, the (fictional) transition from the  magnetically disordered  GS
to a state with magnetic LRO obtained by a continuous increase of spin quantum number $s$
starting from the extreme quantum case $s=1/2$ is always a transition to the state with
$\sqrt{3}\times\sqrt{3}$ LRO, while in Ref.~\onlinecite{zhito_XXZ_2014}
it is suggested that the transition is to the state with $q=0$ LRO.
Moreover, for $s=3/2$ we find $\sqrt{3}\times\sqrt{3}$ LRO for $0.525
< \Delta \le 1$, while in Ref.~\onlinecite{zhito_XXZ_2014} $q=0$  LRO is
suggested for the entire region $0 \le \Delta \le 1$.

We conclude, that the interplay of frustration, quantum fluctuations and
anisotropy  leads to a rich ground-state phase diagram  
of the XXZ spin-$s$ KAFM. Bearing in
mind the numerous investigations of the isotropic Heisenberg KAFM,
see,
e.g.,~Refs.~\onlinecite{Elser1990,Harris1992,chub92,sachdev1992,Reimers1993,
henley1995,Waldtmann1998,Capponi2004,Singh2007,cepas2008,Sindzingre2009,Bishop2010,Evenbly2010,Yan2011,lauchli2011,
lee2011,nakano2011,goetze2011,schollwoeck2012,
Li2012,becca2013,bruce2014,Ioannis2013,zhito_XXZ_2014,XXZ_s12_2014,XXZ_s12_2014_a,cepas2011,Lauchli_s1_2014,Weichselbaum_s1_2014,satoshi_s1_2014,Weichselbaum_s1_2014a} 
and references
therein, the anisotropic model provides a challenging playground to apply
the toolbox of frustrated quantum magnetism on this so far little
investigated problem.

\section*{Acknowledgments}
The authors thank A. L. Chernyshev and M. E. Zhitomirsky
for valuable discussion and also for providing their data used in our Fig.~\ref{fig3}.


\begin{thebibliography}{99}
\bibitem{SUBn-n} Since for $s=1/2$  typically the notation LSUB$n$ is used, 
it is in order to mention here that for $s=1/2$ the LSUB$n$ scheme is identical 
to the SUB$n$-$n$ scheme.
\end{thebibliography}
\end{document}